\documentstyle[11pt]{article}

\input epsfig.sty

\def\be{\begin{equation}}
\def\ee{\end{equation}}
\def\ba{\begin{eqnarray}}
\def\ea{\end{eqnarray}}

\def\bbox{\stackrel{-}{\Box}}
\def\tbox{\stackrel{\sim}{\Box}}
\def\teps{{\widetilde{\epsilon}}^{\mu\nu\lambda\kappa}}
\def\tnabla{\widetilde{\nabla}}
\def\trho{\widetilde{\rho}}
\def\ta{\widetilde{a}}
\def\tb{\widetilde{b}}
\def\tg{\widetilde{g}}
\def\tH{\widetilde{H}}
\def\tp{\widetilde{p}}
\def\tq{\widetilde{q}}
\def\tR{\widetilde{R}}
\def\ts{\widetilde{s}}
\def\tt{\widetilde{t}}
\def\tT{\widetilde{T}}
\def\tTH{\ ^{(H)}\widetilde{T}}
\def\tTV{\ ^{(V)}\widetilde{T}}
\def\tTphi{\ ^{(\phi)}\widetilde{T}}
\def\bara{\bar{a}}
\def\bg{\bar{g}}
\def\half{\frac{1}{2}}

\def\l{\label}

\def\r{\ref}
\def\dh{\dot{h}}

\begin{document}
  \begin{titlepage}
\renewcommand{\thefootnote}{\alph{footnote}}
  \begin{flushright}
SUSX-TH-94/3-7\\
SUSSEX-AST-94/6-2\\
hep-th/9406216\\
  \end{flushright}
  \begin{center}
\vspace{1 in}
\Large
{\bf Low energy effective string cosmology\\}
\vspace{0.2 in}
\normalsize
\large{E.~J.~Copeland\footnote{edmundjc@central.susx.ac.uk},
Amitabha Lahiri\footnote{a.lahiri@central.susx.ac.uk}
\& David Wands\footnote{dwands@star.susx.ac.uk}} \\
\normalsize
\vspace{0.2 in}
{\em  School of Mathematical \& Physical Sciences, \\
University of Sussex, \\ Brighton BN1 9QH.\\U.~K.}\\
\end{center}
\vspace{0.2 in}
\begin{abstract}
We give the general analytic solutions derived from the low energy string
effective action for four dimensional Friedmann-Robertson-Walker models
with dilaton and antisymmetric tensor field, considering both long and short
wavelength modes of the $H$-field. The presence of a homogeneous $H$-field
significantly modifies the evolution of the scale factor and dilaton.
In particular
it places a lower bound on the allowed value of the dilaton. The scale factor
also has a lower bound but our solutions remain singular as they all contain
regions where the spacetime curvature diverges signalling a breakdown in the
validity of the effective action. We extend our results to the simplest
Bianchi I metric in higher dimensions with only two scale factors. We again
give the general analytic solutions for long and short wavelength modes
for the $H$ field restricted to the three dimensional space, which produces an
anisotropic expansion. In the case of $H$ field radiation (wavelengths within
the Hubble length) we obtain the usual four dimensional radiation dominated FRW
model as the unique late time attractor.
\end{abstract}
\vspace{0.2 in}

\end{titlepage}
\renewcommand{\thefootnote}{\arabic{footnote}}
\renewcommand{\theequation}{\thesection.\arabic{equation}}
\setcounter{footnote}{0}
\setcounter{equation}{0}

\section{Introduction}
\setcounter{equation}{0}
\l{sectINTRO}

String inspired cosmology is currently attracting a great deal of
attention. The  most favored starting point in any analysis is the low
energy string  effective action from which the lowest order string
beta-function equations  can be derived \cite{effaction}. These
equations, for the closed string, consist of three long range fields,
the dilaton $\phi$, the Kalb-Ramond field strength $H_{\mu\nu\lambda}$,
and the graviton, all arising out of the massless excitation of the
string. In addition there is a constant related to the central  charge
of the string theory which vanishes in the critical number of
dimensions  10 or 26. The fact that only the massless excited state is
used  suggests that the effective action is not a valid description for
probing the  highest energies associated with string theory. However,
we may hope that through  the beta-function equations we are
investigating physics associated with  events from say the string scale
down to the GUT scale. Such an approach has  already been adopted by a
number of authors \cite{effaction,anton,bailin,liddle 89,ven
91,campbell,tse+Vaf,particular,Tseytlin 92,Gas+Ven 93b,Gas+Ven
93a,Gol+Per 93}.

Veneziano and his collaborators have emphasized the  possible
importance of the duality symmetry which  characterizes the equations
of string cosmology  \cite{ven 91,Gas+Ven 93a,Bru+Ven 94}. They point
out the possibility that  inflation can occur without  relying on a
potential energy density. Rather, the duality transforms, which relate
the dilaton and the metric, lead to a possible  inflationary mechanism.
In \cite{Bru+Ven 94} it is claimed that the  inclusion of the $H$-field
does not seem to change the underlying properties  of these
duality-related cosmologies, although they also point out  possible
problems with exiting inflation in these models. Cosmological
solutions with a dilaton and a non-trivial $H$-field have been obtained
by  Tseytlin\cite{Tseytlin 92}, for curved maximally symmetric spaces
in string  theory, with a non-zero central charge deficit. (See
\cite{Tseytlin 92} for  details and complete references).

In \cite{Gol+Per 93}, the authors study the outcome of string-dominated
 cosmology in a four dimensional Friedmann-Robertson-Walker (FRW)
spacetime including a homogeneous  $H$-field ($H_{\mu\nu\lambda}$) as
well as a non-zero critical charge deficit $V$  (some of the particular
solutions were previously obtained  by Tseytlin \cite{particular}).
Their results are based on a phase-plane  analysis. In this paper, we
present the general analytic solutions, including both long and short
wavelength solutions for the $H$-field, but setting $V=0$, and then
show how this can be extended to simple anisotropic models in higher
dimensions.

In section 2 we introduce the low  energy equations of motion in the
string frame and  demonstrate how they can be related via simple
conformal transforms to  equations in other frames.  Section 3
concentrates on four  dimensional FRW models and describes the complete
solution for a  homogeneous $H$-field, reproducing where applicable
previous solutions  found in the literature \cite{Tseytlin 92,Gol+Per
93}. We go on to  describe radiation solutions in which the $H$-field
has a spatial dependence on small scales (i.e.~wavelengths much smaller
than the Hubble length).  Because the metric (and the dilaton) is
homogeneous and isotropic we require the  $H$-field energy-momentum
tensor to be homogeneous and isotropic on average,  and demonstrate how
at late times we recover the general relativistic  results for
radiation and curvature dominated models.  In section 4 we consider the
 influence of homogeneous and radiation solutions of the  $H$-field in
$D=4+n$ cosmologies where the metric tensor is decomposed into the
direct product of a four dimensional FRW metric and an $n$ dimensional
metric.  We give general analytic solutions for the simplest version of
the $D$ dimensional Bianchi I models where there are just two scale
factors. The results indicate how the $H$-field can produce an
anisotropic expansion leading to one  scale expanding whilst the other
contracts.  Finally in section 5 we summarize our main results.


\section{String action}
\setcounter{equation}{0}
\l{sectACTION}

We shall take as our starting point the low energy $D$-dimensional string
effective action \cite{effaction}
\be
S = \frac{1}{2\kappa_D^2} \int d^Dx \sqrt{-g} e^{-\phi} \left[
	R + \left( \nabla\phi \right)^2 - V
	- \frac{1}{12} H^2 \right]
	+  \int d^Dx \sqrt{-g} {\cal L}_{\rm matter} \; ,
\l{SACT}
\ee
where $\kappa_D^2=8\pi G_D$ and we adopt the sign conventions denoted
(+++) by Misner, Thorne and Wheeler \cite{MTW}. $\phi$ is the dilaton
field determining the strength of the gravitational coupling and
$H^2=H_{\mu\nu\lambda} H^{\mu\nu\lambda}$ where
$H_{\mu\nu\lambda}=\partial_{[\mu}B_{\nu\lambda]}$.

The variation of this action with respect to the $g_{\mu\nu}$,
$B_{\mu\nu}$ and $\phi$, respectively, yields the field equations
\ba
R_{\mu}^{\nu} - \frac{1}{2} g_{\mu}^{\nu} R & = &
	\kappa_D^2 e^{\phi} T_{\mu}^{\nu}
	+ \frac{1}{12} \left( 3H_{\mu\lambda\kappa}H^{\nu\lambda\kappa}
			- \frac{1}{2} g_{\mu}^{\nu} H^2 \right)
	- \frac{1}{2} g_{\mu}^{\nu} V \nonumber \\
& & \quad - \frac{1}{2} g_{\mu}^{\nu} \left( \nabla\phi \right)^2
	+ \left( g_{\mu}^{\nu} g^{\lambda\kappa}
			-g_{\mu}^{\lambda} g^{\nu\kappa} \right)
				\nabla_\lambda\nabla_\kappa\phi
\; , \l{SGMU} \\
\nabla_{\mu} \left( e^{-\phi} H^{\mu\nu\lambda} \right) & = & 0 \; ,
\label{Heom} \l{SH} \\
2\Box\phi + R - \left(\nabla\phi\right)^2 - V - \frac{1}{12}H^2 & = & 0 \; ,
\l{SDIL}
\ea
where $T_{\mu\nu}$ is the energy-momentum tensor derived from the matter
Lagrangian.

The effect of certain types of ``stringy matter'' has been considered
elsewhere in the literature \cite{Gas+Ven 93a}. Specific schemes of
compactification, not to mention the chosen gauge symmetries of the
theory, will determine the behavior (and number) of both bosonic and
fermionic matter fields. It is not inconceivable that in some
`matter-dominated era' of the stringy epoch of the universe these
matter fields will play a part in determining the cosmological
evolution. In particular, the antisymmetric tensor may be considered a
matter field, and as we shall see it plays a significant role in string
cosmology. In favor of considering the possible effects of this field
on the cosmology we shall ignore all other contributions from the
matter Lagrangian.

The charge deficit $V$ is a constant proportional to $D-26$ for the
bosonic string and $D-10$ in the heterotic or superstring. We will set
$V=0$ in our analysis. This may well be necessary for a consistent
theory either by choosing the appropriate number of spatial dimensions
or due to cancellation with contributions from other matter fields.
Even for non-zero $V$ this should be an increasingly good approximation
at early times if the curvature and/or kinetic energy densities are
large, $V\ll R, (\nabla\phi)^2, H^2$, but we would need to consider its
effect at late times in an expanding universe.

\subsection{Conformal frames}
\l{ssectCONF}

These field equations are similar to those found in Brans-Dicke
gravity\cite{BDgravity} with the Brans-Dicke parameter $\omega=-1$.
This is  only strictly true in the absence of the $H$-field as in
Brans-Dicke gravity it is assumed that the energy-momentum tensor of
all fields (other than the Brans-Dicke field, $\Phi\equiv e^{-\phi}$)
are minimally coupled to the metric $g_{\mu\nu}$. While the
energy-momentum tensor of other matter fields are assumed to be
conserved with respect to this metric (the string metric), so that
$\nabla_{\mu}T^{\mu\nu}=0$, we cannot define an energy-momentum tensor
solely in terms of the $H$-field and the string metric which is
conserved independently of the dilaton. This is just a consequence of
the equation of motion for $H$ (Eq.~(\ref{SH})) which has an explicit
dependence upon $e^{\phi}$.

$H_{\mu\nu\lambda}$ is only minimally coupled in the conformally related
metric
\be
\bg_{\mu\nu} = \exp\left(\frac{2\phi}{6-D}\right) g_{\mu\nu} \; ,
\l{BMET}
\ee
which we might call the B-metric, in which we find\footnote
{Note that the 3-form $H_{\mu\nu\lambda}$ has a conformally invariant
definition in terms of the potential $B_{\mu\nu}$, whereas its
covariant form has indices raised by a particular metric.}
\be
\bar{\nabla}_{\mu} \bar{H}^{\mu\nu\lambda} = 0 \; .
\l{BH}
\ee

Another particularly useful metric to introduce is the Einstein
metric\cite{Conformal}
\be
\tg_{\mu\nu} = \exp\left(-\frac{2\phi}{D-2}\right) g_{\mu\nu} \; .
\l{EMET}
\ee
In this frame the action appears simply as the Einstein-Hilbert action of
general relativity in $D$-dimensions, while the dilaton appears simply as
a matter field, albeit one interacting with the other matter fields,
\ba
S & = & \frac{1}{2\kappa_D^2} \int d^Dx \sqrt{-\tg} \left[
	\tR - \frac{1}{D-2} \left( \tnabla\phi \right)^2
	- V \exp\left(\frac{2\phi}{D-2}\right)
	- \frac{1}{12} \exp\left(-\frac{4\phi}{D-2}\right) \tH^2 \right]
\nonumber \\
& & \hspace{1in}
	+ \int d^Dx \sqrt{-\tg}
		\exp\left(\frac{D\phi}{D-2}\right) {\cal L}_{\rm matter} \; .
\l{EACT}
\ea
The corresponding field equations are then those for interacting fields in
general relativity;
\ba
\tR_{\mu\nu} - \frac{1}{2} \tg_{\mu\nu} \tR
= \kappa_D^2 \left( \tT_{\mu\nu} + \tTH_{\mu\nu}
			+ \tTphi_{\mu\nu} + \tTV_{\mu\nu} \right)\; ,
\l{EGMU} \label{einstein} \\
\tnabla_{\mu} \left( \exp\left(-\frac{4\phi}{D-2}\right)
		\tH^{\mu\nu\lambda} \right) = 0 \; ,
\l{EH} \l{einsteinHeom} \\
\tbox\phi - V \exp\left(\frac{2\phi}{D-2}\right)
	+ \frac{1}{6} \exp\left(-\frac{4\phi}{D-2}\right) \tH^2 = 0
\l{EDIL} \l{einsteinphieom}
\; ,
\ea
where the terms on the right-hand side of the Einstein
equations correspond to
\ba
\tT_{\mu}^{\nu} & = & \exp\left(\frac{D\phi}{D-2}\right) T_{\mu}^{\nu}
\; , \l{ETMU} \\
\kappa_D^2 \tTphi_{\mu}^{\nu} & = & \frac{1}{D-2}
	\left( \tg_{\mu}^{\lambda} \tg^{\nu\kappa}
		- \half \tg_{\mu}^{\nu} \tg^{\lambda\kappa} \right)
	\phi_{,\lambda} \phi_{,\kappa} \; , \l{ETDIL} \\
\kappa_D^2
 \tTH_{\mu}^{\nu} & = & \frac{1}{12} \exp\left(-\frac{4\phi}{D-2}\right)
	\left( 3\tH_{\mu\lambda\kappa}\tH^{\nu\lambda\kappa}
		-\frac{1}{2}\tg_{\mu}^{\nu}\tH^2 \right)
\; , \l{ETH} \\
\kappa_D^2
 \tTV_{\mu}^{\nu} & = & - \frac{1}{2} V \exp \left(\frac{2\phi}{D-2}\right)
  \tg_\mu^\nu
\ea
the energy-momentum tensors for the matter, dilaton and $H$-fields and
potential $V$ respectively. The total energy-momentum must be conserved
of course by the Ricci identity, but there are interactions between
these four components. Henceforth, as remarked earlier, we shall set
$T_{\mu\nu}$ and $V$ to be zero.

\section{Isotropic D=4 solutions}
\setcounter{equation}{0}
\label{sectISO}

Firstly we will consider the behavior of $4$-dimensional homogeneous and
isotropic cosmologies for which the most general metric is the
Friedmann-Robertson-Walker metric
\ba
ds^2 & = & -dt^2 + a^2(t) \left( \frac{dr^2}{1-kr^2} + r^2d\Omega^2 \right)
\; , \\
	& = & a^2(\eta) \left( -d\eta^2 +
			\frac{dr^2}{1-kr^2} + r^2d\Omega^2 \right) \; ,
\ea
in terms of the conformally invariant time coordinate, $\eta$, with $k= +1, 0,
-1$
for spatially closed, flat or open models respectively. Just as we take the
metric to be homogeneous we shall also assume that the dilaton has no spatial
dependence, $\phi=\phi(\eta)$.

With $D=4$ the conformal transform to the Einstein frame gives a
rescaled scale factor $\ta=e^{-\phi/2}a$. The metric field equations
are simplest in terms of the Einstein metric where we have the familiar
constraint equation (the $00$-component of Eq.~(\ref{einstein}) with
$V=0$ and $\tT_\mu^\nu=0$)
\be
-3 \left( \frac{\ta'^2}{\ta^4} + \frac{k}{\ta^2} \right) \ = \
		\kappa^2 \left( \tTH_0^0 + \tTphi_0^0 \right) \; ,\\
\label{frwconstraint}
\ee
where prime denotes differentiation with respect to $\eta$.

In $4$-dimensions the $H$-field equation of motion, Eq.~(\ref{einsteinHeom}),
is solved by the Ansatz,
\be
\tH^{\mu\nu\lambda} = e^{2\phi} \teps h_{,\kappa} \; ,
\ee
where $\teps$ is the antisymmetric 4-form (obeying $\tnabla_{\rho}\teps=0$)
and the integrability condition,
$\partial_{[\mu}\tH_{\nu\lambda\kappa]}=0$, becomes the new equation of motion
for $h$
\be
\tbox\,h + 2 \tnabla^{\mu}\phi \tnabla_{\mu}h = 0 \; .
\label{einsteinheom}
\ee
Thus $h$ evolves as a massless scalar field coupled to the dilaton (except
in the B-frame where $\bbox\,h=0$).
The same interaction appears in the dilaton equation of motion
(Eq.~(\ref{einsteinphieom}), with $V=0$) as
\be
\tbox\,\phi = e^{2\phi} \left(\tnabla h\right)^2 \; .
\ee

Thus we have two interacting scalar fields whose energy-momentum tensors
are given by
\ba
\kappa^2 \tTphi_{\mu}^{\nu} & = & \frac{1}{2}
	\left( \tg_{\mu}^{\lambda} \tg^{\nu\kappa}
		- \half \tg_{\mu}^{\nu} \tg^{\lambda\kappa} \right)
		\phi_{,\lambda}\phi_{,\kappa} \; , \\
\kappa^2 \tTH_{\mu}^{\nu} & = & \frac{1}{2}
	\left( \tg_{\mu}^{\lambda} \tg^{\nu\kappa}
		- \half \tg_{\mu}^{\nu} \tg^{\lambda\kappa} \right)
		e^{2\phi} h_{,\lambda} h_{,\kappa} \; .
\ea

These equations simplify considerably in two cases.

\subsection{Homogeneous solution $h(\eta)$}
\l{ssectISOhom}

Thus far we have allowed for the possibility of a spatial dependence of
the $H$-field. However as we have already restricted ourselves to
considering a homogeneous metric and dilaton field this will only be
consistent with choosing a source that is at least homogeneous on
average. Indeed as far as we are aware, the only case that has been
considered to date\cite{Gol+Per 93} is that of a strictly isotropic
$H$-field, where $h=h(\eta)$, and thus $\tH_{0\mu\nu}=0$.

In this case the equation of motion for the $H$-field
(Eq.~(\ref{einsteinheom})) becomes
\be
h'' + \left( 2\frac{\ta'}{\ta} + 2\phi' \right) h' = 0 \; ,
\ee
which is easily integrated to give $e^{2\phi}\ta^2h'=\pm L$ where $L$
is a non-negative constant. It is this kinetic energy density of the
$H$-field that drives the dilaton. Note that back in the string frame
this solution corresponds to $H^2=6L^2/a^6$. Thus it will dominate any
charge deficit $V$ in the dilaton equation of motion, Eq.~(\ref{SDIL}),
as $a\to0$.

Because both $h$ and $\phi$ are
functions only of time, we can define a new scalar field $\psi(t)$ where
\be
d\psi^2 = d\phi^2 + e^{2\phi} dh^2 \; .
\ee
The energy-momentum tensor in the Einstein frame is then simply that for this
single minimally coupled field
\be \kappa^2 \left( \tTphi_{\mu}^{\nu} + \tTH_{\mu}^{\nu} \right) =
\half
	\left( \tg_{\mu}^{\lambda} \tg^{\nu\kappa}
		- \half \tg_{\mu}^{\nu} \tg^{\lambda\kappa} \right)
	\psi_{,\lambda} \psi_{,\kappa} \; .
\ee
The equation of motion for a homogeneous minimally coupled field is
\be
\psi'' + 2\frac{\ta'}{\ta}\psi' = 0 \; ,
\ee
which can be integrated to give
$\ta^2\psi'=\pm K$, where $K$ is a positive constant. (We will not consider
the trivial case where $K=0$.) Thus in the Einstein
frame we have an isotropic perfect stiff fluid whose energy density
$\kappa^2\trho\equiv -\kappa^2(\tTphi_0^0+\tTH_0^0)=K^2/4\ta^6$.

The constraint Eq.(\ref{frwconstraint}) for the Einstein scale factor, $\ta$,
can then be integrated to give\cite{Mimoso+Wands 94}
\be
\ta^2 = \frac{K}{\sqrt{3}} \frac{\tau}{1+k\tau^2} \; ,
\ee
where we define
\be
\tau(\eta) \ = \ \left\{
\begin{array}{cc}
| \tan (\eta-\eta_0) |  	& {\rm for} \ k=+1 \; , \\
| \eta-\eta_0 | 		& {\rm for} \ k=0 \; , \\
| \tanh (\eta-\eta_0) | 	& {\rm for} \ k=-1 \; .
\end{array} \right.
\label{edeftau}
\ee
We emphasize that in the Einstein frame the scale factor evolves in a wholly
unremarkable fashion\cite{Gas+Ven 93b}. We have a singularity at $\eta=\eta_0$
with $\ta=0$ and the models expand
away from it for $\eta>\eta_0$ or collapse towards it for
$\eta<\eta_0$. Only closed models can turn around, and these recollapse at
$\eta=\eta_0\pm\pi/2$. There are no bounce solutions. Notice also that the
behavior of the Einstein scale factor is independent of $L$, and thus is the
same in vacuum, i.e.~without the presence of the $H$-field, as it is with the
$H$-field.

Combining the first integrals for $\psi$ and the $H$-field with the definition
of $d\psi$ we also have
\be
\phi '^2 = \frac{K^2-e^{-2\phi}L^2}{\ta^4} \; .
\ee
This too can be integrated to give
\ba
e^{\phi}  = \left\{
\begin{array}{cc}
\left( \frac{\tau}{\tau_*} \right)^{\pm\sqrt{3}}
& {\rm for} \ L=0 \; , \\
\frac{L}{2K} \ \left(
		\left(\frac{\tau_*}{\tau}\right)^{\sqrt{3}}
		+ \left(\frac{\tau}{\tau_*}\right)^{\sqrt{3}} \right)
& {\rm for} \ L\neq0 \; ,
\end{array} \right.
\l{ephitauhom}
\ea
where $\tau_*$ is an integration constant. Note that the evolution in the
presence of the antisymmetric tensor field is quite distinct from the vacuum
behavior. In particular there is a lower bound on the dilaton, $e^{2\phi}\geq
K^2/L^2$. By contrast, there are two distinct branches in vacuum where the
dilaton is either monotonically increasing or decreasing.

We can use this to recover the scale factor in the string frame
\ba
a^2 = \left\{
\begin{array}{cc}
\frac{K\tau_*}{\sqrt{3}} \
	\left({\tau \over  \tau_*}\right)^{1\pm\sqrt{3}}
	\left(1+k\tau^2\right)^{-1}
& {\rm for} \ L=0
\; , \\
\frac{L\tau_*}{2\sqrt{3}} \
	\left( \left({\tau \over \tau_*}\right)^{-\sqrt{3}+1}
		+ \left({\tau \over \tau_*}\right)^{\sqrt{3}+1} \right)
	\left(1+k\tau^2\right)^{-1}
& {\rm for} \ L\neq0 \; .
\end{array} \right.
\l{atauhom}
\ea
The evolution of the scale factor and dilaton in different cases is shown in
Figs.~(1--4).
The singular behavior of the conformal factor, $e^{\phi}$, at the initial
singularity in the Einstein frame produces cosmologies which have a
minimum non-zero value for the scale factor
in the string frame. In the presence of the $H$-field, $a$ diverges both as
$\eta\to\pm\infty$ (or as $\eta\to\eta_0\pm\pi/2$ for $k>0$) and as
$\eta\to\eta_0$, so all models ``bounce'', although they are still singular in
the sense that the Ricci curvature diverges.

\vbox{\leftline{\hbox{\epsfig{width=3.3in,file=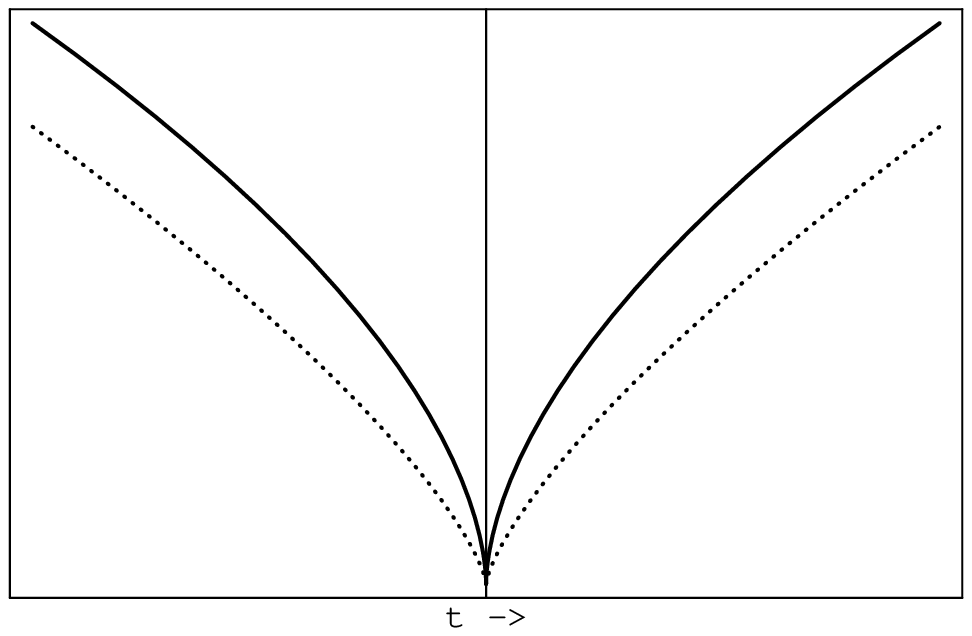}
\epsfig{width=3.3in,file=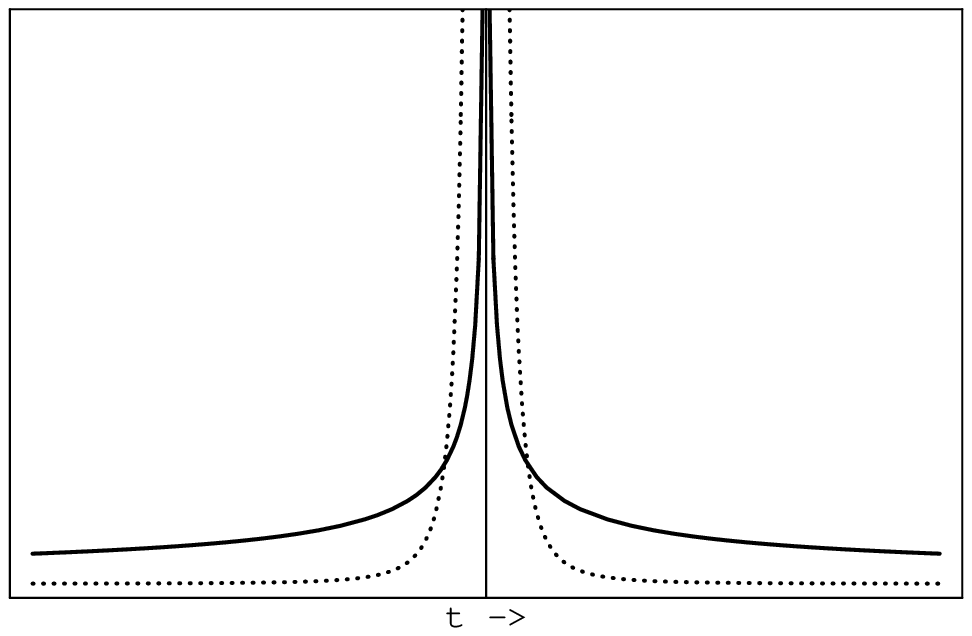}}}\break
{\small {\bf Fig.~1:} The decelerated branches (left) and the
accelerated branches of the scale factor $a$ (solid line) and dilaton
$e^{\phi}$ (dotted) in a spatially flat FRW universe with vanishing
(or constant) $h$.}}

\bigskip

Note that the vacuum solutions again exhibit two distinct branches
corresponding to $\phi$ and $a$ either monotonically increasing or
decreasing. The $k=0$ models correspond simply to power-law solutions
shown in Fig.~1. We can write the solutions in terms of the
proper time in the string frame by integrating $dt=ad\eta$. The two
solutions are then
\begin{itemize}
\item $e^{\phi}\propto|t-t_0|^{+\sqrt{3}-1}$ and
$a\propto |t-t_0|^{+1/\sqrt{3}}$;
\item $e^{\phi}\propto|t-t_0|^{-\sqrt{3}-1}$ and
$a\propto |t-t_0|^{-1/\sqrt{3}}$.
\end{itemize}
These two branches, corresponding to a decelerated or accelerated
scale factor, are related by the duality transformation\cite{Mei+Ven
91} $a\to1/a$ and $e^{\phi}\to e^{\phi}/a^6$.

For $L\neq0$ the $k=0$ solution corresponds to the power-law vacuum
solutions at early and late times and we find that it smoothly
interpolates between the accelerated and decelerated branches. (See
Fig.~2.)  In some sense then it could be described as a ``self-dual''
solution.
\begin{itemize}
\item as $\eta\to\eta_0$, we have $e^{\phi}\propto|t-t_0|^{-\sqrt{3}-1}$ and
$a\propto |t-t_0|^{-1/\sqrt{3}}$ as $t\to t_0$;
\item as $\eta\to\pm\infty$, we have $e^{\phi}\propto|t|^{\sqrt{3}-1}$ and
$a\propto |t|^{1/\sqrt{3}}$ as $t\to\pm\infty$.
\end{itemize}
This is precisely the behavior found by Goldwirth and
Perry\cite{Gol+Per 93} in their phase-plane analysis. The vacuum solutions
correspond to particular solutions (here corresponding to the limiting
behavior where $\tau_*$ is either infinite or zero) found
previously\cite{particular}.

The solutions in spatially curved models approach the flat space
results only near $\eta_0$. At late times $\tau\to1$ in open ($k=-1$)
models (Fig.~3) and thus the dilaton becomes frozen-in at a fixed value
as the curvature dominates the evolution and and we approach the
Einstein result. In closed models (Fig.~4) where $\tau\to\infty$ as
$\eta-\eta_0\to\pm\pi/2$, the scale factor diverges in a finite proper
time. Thus although these models bounce, and therefore must undergo a
period of inflation ($\ddot{a}>0$) in the string frame, they still
become curvature dominated at late times.

\vbox{\centerline{\epsfig{width=3.3in,file=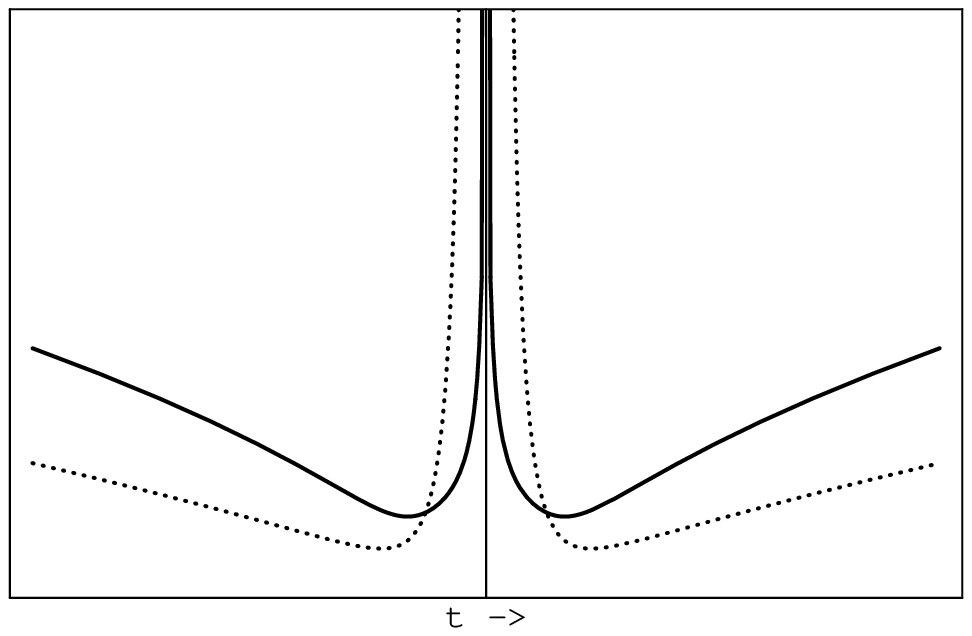}}
\noindent\small
{\bf Fig. 2:}
Scale factor $a$ (solid line) and dilaton $e^{\phi}$ (dotted) in a
spatially flat FRW universe with homogeneous $h$.}

\vbox{\centerline{\epsfig{width=3.3in,file=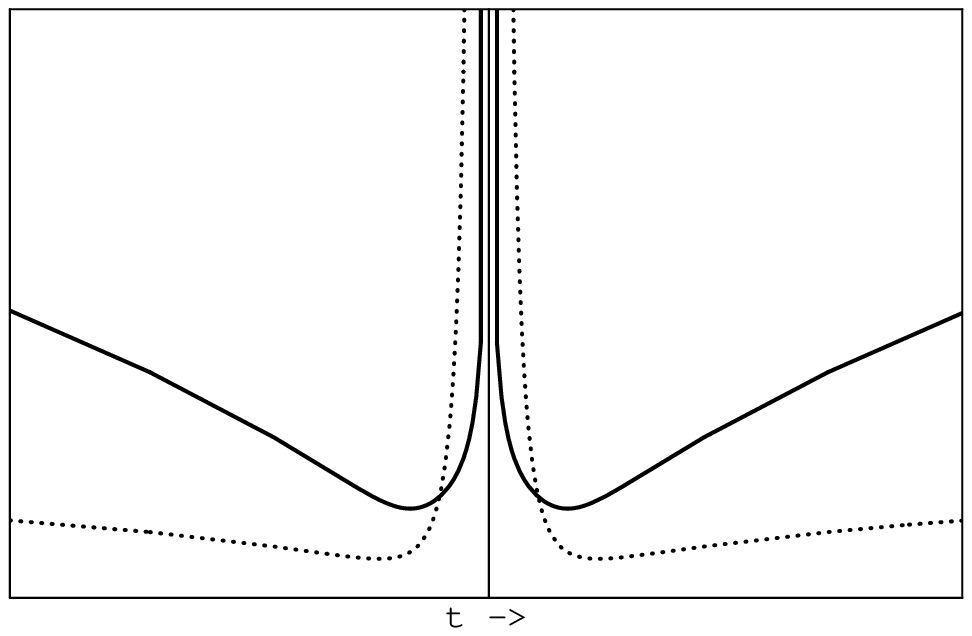}}
\noindent\small
{\bf Fig. 3:}
Scale factor $a$ (solid line) and dilaton $e^{\phi}$ (dotted) in an
open ($k=-1$) FRW universe with homogeneous $h$.}

\vbox{\centerline{\epsfig{width=3.3in,file=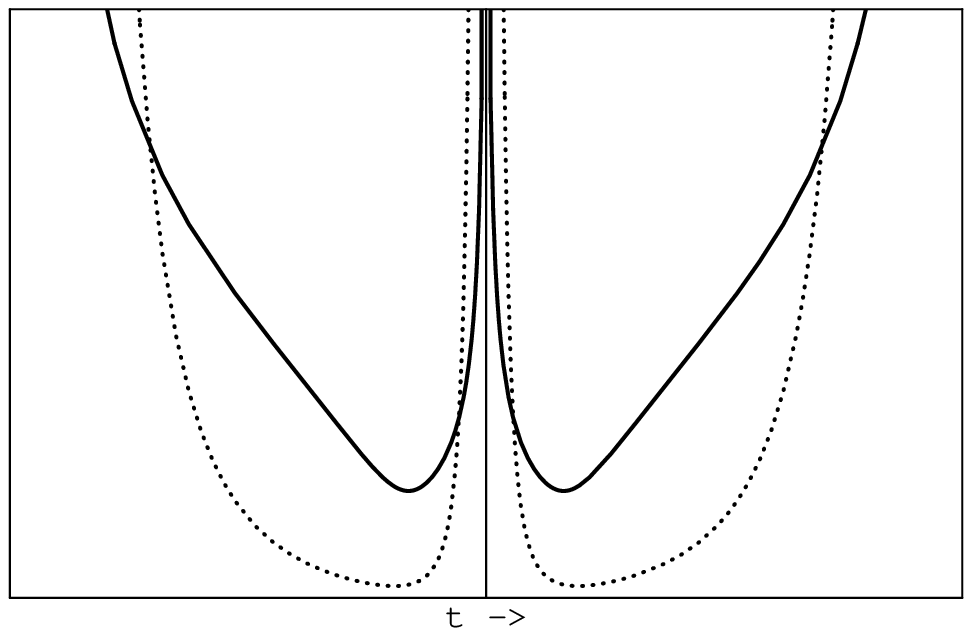}}
\noindent\small
{\bf Fig. 4:}
Scale factor $a$ (solid line) and dilaton $e^{\phi}$ (dotted) in a
closed ($k=+1$) FRW universe with homogeneous $h$.}

\subsection{Radiation solution, $(\nabla h)^2=0$}
\l{ssectISOrad}

It is also possible to consider cases where the $H$-field does have a
spatial dependence. Because our metric (and dilaton) is homogeneous and
isotropic we will require that the $H$-field energy-momentum tensor is
also homogeneous and isotropic on average. It is natural (in flat
space) to decompose any field into Fourier modes,
$h=\sum_qh_q(\eta)\exp(iq_ix^i)$ where $q_i$ is a spatial comoving
three-vector. We see then that the preceding case corresponds to the
long wavelength mode where $q^2=\sum_i q_i^2\to0$. The other case in
which we can solve the equation of motion is where $q\to\infty$ where
effects of spacetime curvature can be neglected and the usual Minkowski
spacetime result holds.

Specifically we find that, in the B-frame, for
$q^2\gg \bara''/\bara ,\ k/\bara^2$,
the equation of motion for $h_q(\eta)$ reduces to
\be
(\bara h_q)'' + q^2 (\bara h_q) \simeq 0 \; .\\
\ee
Thus, for a single short wavelength mode, in the limit $q\to\infty$, we
have $h_q= m_q e^{iq\eta}/\bara$ where $m_q$ is a constant. This
corresponds to an energy-momentum tensor for the $H$-field in the
Einstein frame
\be
\kappa^2 \tTH_{\mu}^{\nu} \ = \ q_{\mu}q^{\nu} \frac{m_q^2}{\ta^2} \; ,\\
\ee
where $q_{\mu}$ is a null 4-vector with $q_0=q$.
Clearly this is not isotropic for a single wave-vector, $q_{\mu}$, but if we
consider an isotropic distribution of wave-vectors we find
\ba
- \kappa^2 \langle \ \tTH_0^0 \ \rangle
 & = & 3 \kappa^2 \langle \ \tTH_i^i \ \rangle \quad {\rm (no \ sum)} \; ,\\
 & = & \frac{M}{\ta^4} \; ,
\ea
the usual result for radiation in a FRW universe, where
$M\equiv\int m_q^2q^2dq$.

Note that the fluid is trace-free, and thus conformally invariant. This
means that the energy-momentum of the radiation is conserved in all
conformal frames and so the $H$-field is decoupled from the dilaton
which appears as a minimally coupled scalar field in the Einstein
frame.  Thus the first integral of its equation of motion, for a
homogeneous $\phi(\eta)$, gives
\be
\phi' = \frac{A}{\ta^2} \; .
\label{ephi'rad}
\ee
Just like the $\psi$ field for
homogeneous $h(\eta)$, the energy-momentum of the $\phi$ field behaves
like a stiff fluid, with
\ba
- \kappa^2 \tTphi_0^0 & = & \kappa^2 \tTphi_i^i \quad {\rm (no \ sum)} \ \ ,\\
	& = & \frac{K^2}{4\ta^6} \; .
\ea

Once again the constraint Eq.(\ref{frwconstraint}), now for two non-interacting
fluids, one radiation, one stiff, can be integrated\cite{Mimoso+Wands 94} to
give
\be
\ta^2 = \frac{K+M\tau}{\sqrt{3}} \ \frac{\tau}{1+k\tau^2} \; ,
\ee
using the time coordinate defined in Eq.~(\ref{edeftau}). (As in the
homogeneous case, this is just the familiar behavior for a FRW universe
in general relativity with matter obeying the
strong energy condition and thus singular for all
models at $\tau=0$.) This in turn allows
the equation for $\phi'$, Eq.~(\ref{ephi'rad}), to be integrated, yielding
$e^{\phi}$ and thus, via the conformal transformation, the scale factor in the
string frame.
\ba
e^{\phi} & = & \left(\frac{s}{s_*}\right)^{\pm\sqrt{3}} \; ,\\
a^2 & = & \left(\frac{s}{s_*}\right)^{\pm\sqrt{3}} \
		\frac{K+M\tau}{\sqrt{3}} \ \frac{\tau}{1+k\tau^2} \; ,
\ea
where we have introduced
\be
s(\eta) \ = \ \frac{M\tau(\eta)}{K+M\tau(\eta)} \; .
\l{defs}
\ee

Notice how the evolution of the dilaton field is now similar to that in the
vacuum case except it is determined by the function $s(\eta)$ rather than
$\tau(\eta)$. At early times ($\eta\sim\eta_0$) $s$ is proportional to $\tau$
but, unlike the vacuum case, $s\to1$ as $\tau\to\infty$ and so
the field becomes frozen in at late times in the flat model as well as the
open model (where $\tau\to1$ and thus $s\to M/(K+M)$). Thus we recover the
late time general relativistic results for radiation and curvature dominated
models respectively.

In common with the vacuum solutions, there are two distinct branches
(Fig. 5) with the dilaton monotonically increasing from zero (the
decelerated branch) or decreasing from infinity (the accelerated
branch) at $\eta=\eta_0$.  For $M\tau\ll |K|$ the evolution is
essentially identical to that in the vacuum case with two branches
where $a\to\infty$ as $\tau\to0$ when $\phi\to\infty$, but $a\to0$
when $\phi\to0$. Thus we have no ``bounce'' solution interpolating
between the vacuum branches.

\vbox{\leftline{\hbox
{\epsfig{width=3.3in,file=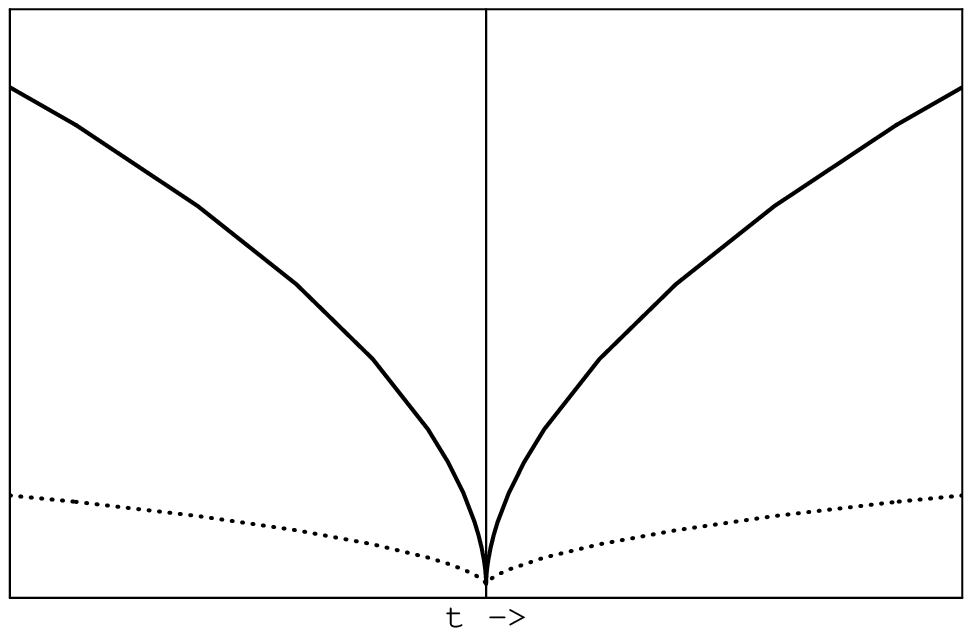}
\epsfig{width=3.3in,file=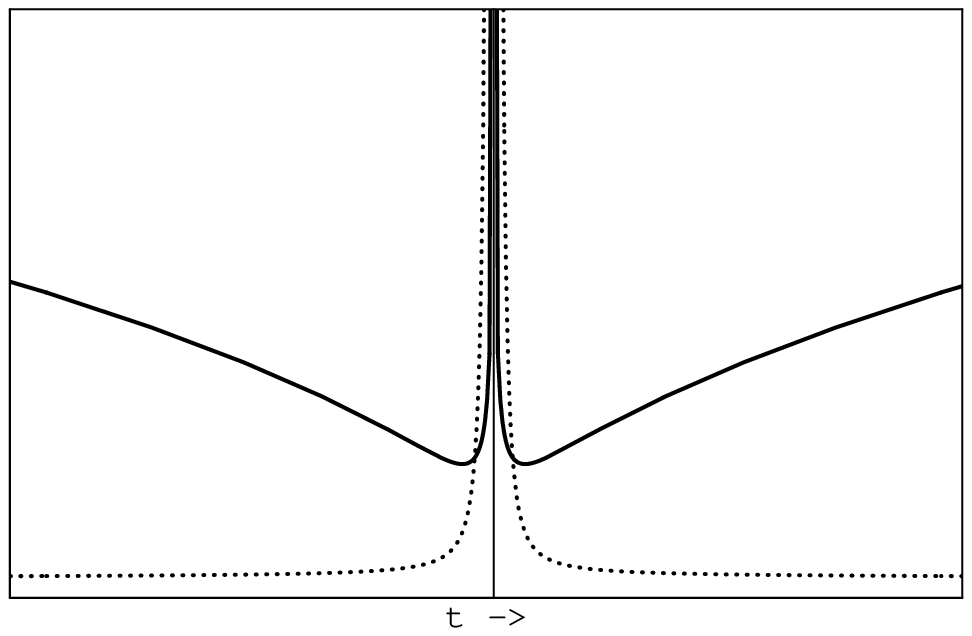}}}\break
{\small
{\bf Fig. 5:}
The decelerated branches (left) and the accelerated branches (right)
of the scale factor $a$ (solid line) and
dilaton $e^{\phi}$ (dotted) in a
spatially flat FRW universe with short wavelength $h$.}}


\section{Anisotropic $D=n+4$ solutions}
\setcounter{equation}{0}
\l{sectANI}

Having investigated the homogeneous and radiation
solutions for the $H$-field in four
dimensions, we return to Eqs.~(\r{EACT})--(\r{ETH}) in the Einstein
frame where we will consider a metric tensor in $D=4+n$ dimensions
that can be
decomposed into the direct product form
\be
d\ts^2 = -d\tt^2 + \ta^2(\tt)dx_idx^i + \tb^2(\tt)dx_Jdx^J \; ,
\l{ELINE}
\ee
where we let $i$ run from $1$ to $3$ and $J$ run from $4$ to $n+3$. The scale
factors $\ta$ and $\tb$ thus refer to the $3$-space and $n$-space split
respectively. We have chosen these spaces to be spatially flat, thus this is a
Bianchi I metric. The procedure we outline here could also be applied to the
general $D$-dimensional Bianchi I metric, but we restrict our discussion here
to a two-scale factor model to avoid introducing too many degrees of freedom.
Eq.~(\r{ELINE}) is of course
related to the original string metric through the conformal transformation
Eq.~(\r{EMET}).

The Einstein evolution equations, Eq.~(\r{EGMU}), for the two scale factors
written in terms of $\dot{\alpha}\equiv(d\ta/d\tt)/\ta$ and
$\dot{\beta}\equiv(d\tb/d\tt)/\tb$ are then
\ba
\ddot{\alpha} + (3\dot{\alpha} + n\dot{\beta})\dot{\alpha}
& = & {\kappa_D^2 \over n+2} \left( \trho + (n-1)\tp - n\tq \right) \; ,
\l{eomalphahom}
\\
\ddot{\beta} + (3\dot{\alpha} + n\dot{\beta})\dot{\beta}
& = & {\kappa_D^2 \over n+2} \left( \trho + 2\tq - 3\tp \right) \; ,
\l{eombetahom}
\ea
plus the constraint equation
\be
3\dot{\alpha}^2 + 3n\dot{\alpha}\dot{\beta} + {n(n-1) \over 2}\dot{\beta}^2
 = \kappa_D^2 \trho \; ,
\l{aniconstraint}
\ee
where $\trho=-\tT_0^0$, $\tp=\tT_i^i$ and $\tq=\tT_J^J$ (no sum over $i$ or
$J$). Note that any isotropic stiff fluid for which $\trho=\tp=\tq$ makes no
contribution to the right-hand-side of the evolution equations, entering only
into the constraint equation.

We will adopt the simplest extension of our $D=4$ Ansatz for the $H$-field,
\be
\tH^{\mu \nu \lambda} = e^{4\phi/(n+2)} \teps h_{,\kappa} \; ,
\l{EHANSATZ}
\ee
where
\be
\teps = \frac{4!}{\sqrt{-\tg}} \delta_{[0}^{\mu} \delta_1^\nu
\delta^\lambda_2 \delta^\lambda_{3]} \; . \nonumber
\ee
Note that whereas this Ansatz included all the solutions in $4$ dimensions,
this represents only one of many degrees of freedom for the $H$-field in the
$D=4+n$ case. An antisymmetric field in the higher-dimensional space
corresponds to a whole range of fields in the $4$-dimensional reduced
theory\cite{salam 82}.
At a classical level we can obtain self-consistent results if these other
fields are set to zero initially, though in a full quantum treatment we
would have to consider precisely how the massive modes of these fields are
excited in the extra dimensions.
Our choice of Ansatz makes any dependence of the function $h$ on the $n$-space
coordinates, $x^J$, irrelevant as this cannot affect $\tH^{\mu\nu\lambda}$ and
so we need consider only $h=h(\tt,x^i)$. A similar approach of demanding the
$H$-field live only in three space dimensions was considered in
\cite{liddle 89} who found numerically that in the Einstein frame
the effect was to drive that space to a large
size while the other spaces remained of order the Planck scale.

Using (\r{EHANSATZ}) the integrability condition
becomes the equation of motion for
$h(\tt,x^i)$;
\be
\sum_{\mu,\nu=0}^3 \tg^{\mu\nu} \partial_{\mu} \left(
	\ta^3 \tb^{-n} e^{-4\phi/(n+2)} h_{,\nu} \right)
= 0 \; .
\ee
We can decompose the field $h$ into its Fourier components,
$h(\tt,x^i)=h_q(\tt)\exp(iq_ix^i)$ so that the equation of motion for the
homogeneous function $h_q$ is
\be
\ddot{h}_q + \left( 3\dot{\alpha} -n\dot{\beta}
	+ {4 \over n+2} \dot{\phi} \right) \dot{h}_q + {q^2 \over \ta^2} h_q
 = 0 \; .
\ee

The final equation of motion is that for the dilaton in the Einstein frame,
Eq.(\ref{einsteinphieom}), driven by
$\tH^2=-6\exp(8\phi/(n+2))\tb^{-2n}(\tnabla h)^2$ giving
\be
\ddot{\phi} + (3\dot{\alpha} + n\dot{\beta})\dot{\phi}
 = - {e^{4\phi/(n+2)} \over \tb^{2n}} \left(\tnabla h\right)^2
\l{eomphi} \; .
\ee

Note that having chosen all the fields to be independent of the
coordinates on the $n$-space we could replace the $D$ dimensional action
by an effective theory written in terms of the four
dimensional part of the metric in the string frame
\be
S_4 = \frac{1}{2\kappa^2} \int d^4x \sqrt{-g_4}\ e^{-\varphi} \left[
	R_4 + \left( \nabla\varphi \right)^2 - n\left(\nabla(\ln b)\right)^2
	- \frac{1}{12} H^2 \right] \; ,
\ee
where the effective $D=4$ dilaton is given by
\be
e^\varphi \equiv {e^\phi \over b^n} \; , \l{defvarphi}
\ee
and the scale factors of the
extra dimensions just act as massless (moduli) fields.
However to emphasize the dynamical evolution of the $n$-space
we shall treat this scale factor on an equal footing with that of the
$3$-space.
Note that the effective $D=4$
Einstein frame would not be the same as the
$D=4+n$ dimensional Einstein frame due to the modified dilaton.

\subsection{Homogeneous solution $h(\tt)$}
\l{ssectANIhom}

The equation of motion for $h$ for long-wavelength modes (where
$q^2\to0$) becomes
\be
\ddot{h} + \left( 3 \dot{\alpha} - n \dot{\beta} + {4 \over n+2} \phi
\right) \dh = 0 \; , \nonumber
\ee
or
\be
\dh = {\pm L \tb^n \over \ta^3} e^{-4\phi/(n+2)} \; ,
\l{EHDOT}
\ee
where $L$ is a positive constant. {}From Eq.~(\r{ETH}) and
making use of Eqs.~(\r{EHANSATZ}) and (\r{EHDOT}) we have
\ba
\tH_{\mu \lambda \kappa}\tH^{\nu \lambda \kappa} & = & 0 \quad {\rm if}\,
\mu ~ {\rm or} ~ \nu = \{0,4,5,..n+3\} \; ,  \nonumber\\
\tH_{i \lambda \kappa}\tH^{j \lambda \kappa} & = & {2L^2 \over \ta^6}
\delta^j_i \quad {\rm for}~i ~ {\rm and} ~ j=\{1,2,3\} \; . \nonumber
\ea
Hence in Eq.~(\r{ETH}) we obtain
\ba
\kappa_D^2 \tTH^0_0 &\equiv& -\kappa_D^2 \trho_H
	= -{L^2 \over 4 \ta^6} e^{-4\phi/(n+2)} \l{ETH0} \; , \\
\kappa_D^2 \tTH^i_i &\equiv& \kappa_D^2 \tp_H
	= {L^2 \over 4 \ta^6} e^{-4\phi/(n+2)} \quad {\rm (no\ sum)}
\l{ETHI} \; , \\
\kappa_D^2 \tTH^J_J &\equiv& \kappa_D^2 \tq_H
	= -{L^2 \over 4 \ta^6} e^{-4\phi/(n+2)} \quad {\rm (no\ sum)}
\l{ETHJ} \; ,
\ea
which means that the $\tH$-field acts like an anisotropic fluid satisfying
\be
\trho_H = \tp_H = -\tq_H \nonumber \; .
\ee
Note that, just as in the $D=4$ case, $H^2$ in the string frame is
inversely proportional to the square of the volume of the 3-space
($H^2\propto a^{-6}$).

There is also a contribution to the energy-momentum from the dilaton field
given by Eq.~(\r{ETDIL}),
\ba
\kappa_D^2 \tTphi^0_0 &\equiv& -\kappa_D^2 \trho_\phi
	= -{\dot{\phi}^2 \over 2(n+2)} \; ,
\l{ETDIL0}\\
\kappa_D^2 \tTphi^i_i &=& \kappa_D^2 \tTphi^J_J
	= {\dot{\phi}^2 \over 2(n+2)} \quad {\rm (no\ sum)} \; ,
\l{ETDILIJ}
\ea
which means that the $\phi$ field acts like an isotropic stiff fluid,
\be
\trho_\phi = \tp_\phi = \tq_\phi \; , \nonumber
\ee
and thus, as remarked in the preceding section, drops out of the evolution
equations for $\alpha$ and $\beta$.

We can substitute the energy-momentum tensors into Eqs.~(\r{EDIL}),
(\r{eomalphahom}) and~(\r{eombetahom})
to obtain the following equations of motion,
\ba
\ddot{\phi} + (3\dot{\alpha} + n\dot{\beta})\dot{\phi} & = &
{L^2 \over \ta^6} e^{-4\phi/(n+2)} \; , \nonumber \\
\ddot{\alpha} + (3\dot{\alpha} + n\dot{\beta})\dot{\alpha} & = &
{n \over 2(n+2)}{L^2 \over \ta^6} e^{-4\phi/(n+2)} \; , \nonumber \\
\ddot{\beta} + (3\dot{\alpha} + n\dot{\beta})\dot{\beta} & = &
-{1 \over (n+2)}{L^2 \over \ta^6} e^{-4\phi/(n+2)} \; , \nonumber
\ea
while the constraint equation becomes
\be
3\dot{\alpha}^2 + 3\dot{\alpha}\dot{\beta} + {n(n-1) \over 2}\dot{\beta}^2 =
{\dot{\phi}^2 \over 2(n+2)} + {L^2 \over 4 \ta^6} e^{-4\phi/(n+2)}
\l{EANICONSTRAINT} \; .
\ee
Notice how the presence of the $H$-field on the right-hand-side of these
evolution equations tends to drive $\phi$ and $\alpha$ in a positive direction,
but drives $\beta$ negative. Thus it produces shear and an anisotropic
expansion.

Introducing a new time coordinate $\xi$ through
\be
d\xi \ \equiv \ \ta^{-3} \tb^{n} e^{-4\phi/(n+2)} d\tt
\l{XI} \; ,
\ee
we obtain first integrals for the equations of motion,
\ba
\phi' &=& {L^2 e^{4\phi/(n+2)} \over \tb^{2n}} \ (\xi + \xi_\phi)
\l{EOMPHI} \; , \\
\alpha' &=& {n \over 2(n+2)} {L^2 e^{4\phi/(n+2)} \over \tb^{2n}} \
(\xi + \xi_\alpha) \; , \l{EOMALPHA}\\
\beta' &=&  -{1 \over (n+2)}
{L^2 e^{4\phi/(n+2)} \over \tb^{2n}} \ (\xi + \xi_\beta) \; , \l{EOMBETA}
\ea
and
\be
h = \pm L(\xi + \xi_h) \; , \nonumber
\ee
where $\phi' \equiv d\phi/d\xi$ etc, and
$\xi_\phi,\xi_\alpha,\xi_\beta$ and $\xi_h$ are constants of
integration. In fact at least one of these is redundant as the origin
of the variable $\xi$ is clearly arbitrary as $\xi$ is only defined by
the differential relation in Eq.(\r{XI}). Henceforth we shall take
$\xi_\beta=0$ so that $\beta'=0$ at $\xi=0$. We can solve for $\beta'$
by differentiating Eq.~(\r{EOMBETA}) and  substituting in for $\phi'$
from Eq.~(\r{EOMPHI}) to obtain
\be
\beta' \equiv {\tb' \over \tb} = {\xi \over
(n+2)\xi^2 + 4\xi_\phi \xi + C} \; ,
\l{BETAPRIME}
\ee
where $C$ is another integration constant related to the others via the
constraint Eq.~(\r{EANICONSTRAINT}).
\be
(n+2)C=2(n+2)\xi_\phi^2 - 3n^2\xi_\alpha^2 \; .
\l{CFIX}
\ee

There is another important constraint which
emerges. {}From Eqs.~(\r{EOMBETA}) and~(\r{BETAPRIME}), by demanding that the
combination $\tb^{2n}e^{-4\phi/(n+2)}$ remains non-negative we obtain
\be
(n+2)\xi^2 + 4\xi_\phi \xi + C \leq 0 \; ,
\l{CCONSTRAIN}
\ee
which means that the allowed range of $\xi$ is bounded by $m_-\leq\xi\leq m_+$,
where $m_{\pm}$ are the roots of the above expression;
\be
m_{\pm} = -{2\xi_\phi \over (n+2)} \pm {\Delta \over n+2} \; ,
\l{MPM}
\ee
and we have introduced $\Delta \equiv [3n^2\xi_\alpha^2 - 2n\xi_\phi^2]^{1/2}$
Clearly solutions only exist for $3n\xi_\alpha^2>2\xi_\phi^2$.

We solve Eq.~(\r{BETAPRIME}) to obtain
\be
\tb(\xi) = \tb_0 (\xi - m_-)^{q_-} (m_+ - \xi)^{q_+} \; ,
\l{BSOLN}
\ee
where $\tb_0$ is a constant and
\be
q_{\pm} = {1 \over 2(n+2)} \left[ 1 \mp {2\xi_\phi \over \Delta} \right] \; .
\l{PPM}
\ee
We obtain similar solutions for $\ta$ and $\phi$ using
Eqs.~(\r{EOMALPHA}), (\r{EOMPHI}) and (\r{BSOLN})
\ba
\ta(\xi) &=& \ta_0 (\xi - m_-)^{p_-} (m_+ - \xi)^{p_+} \l{ASOLN} \; , \\
e^{\phi(\xi)} &=& e^{\phi_0}
(\xi - m_-)^{s_-} (m_+ - \xi)^{s_+} \l{PHISOLN} \; ,
\ea
where $\ta_0$ and $\phi_0$ are constants
(with $\tb_0^ne^{-2\phi_0/(n+2)}=L$ ),
$p_{\pm}$ and $s_{\pm}$ being given by
\ba
p_{\pm} &=& - {n\over 4(n+2)} \left[ 1
	\mp {2\xi_\phi - (n+2)\xi_\alpha \over \Delta} \right] \l{QPM} \; ,\\
s_{\pm} &=& -\half \left[ 1 \pm {n\xi_\phi \over \Delta}
\right] \l{RPM} \; .
\ea

It is clear from Eqs.~(\r{BSOLN}), (\r{ASOLN}) and (\r{PHISOLN}) that
the behavior of $\ta,~\tb$ and $e^{\phi}$ is similar in each case,  the
differences emerging in the exponents of the $(\xi - m_{\pm})$. We need
to know how these solutions appear in the string frame as this is where
 the original theory has emerged from. This is straightforward to do.
Using  Eqs.~(\r{EMET}) we see that the scale factors in the string
frame are given by
\ba
a(\xi) &=& a_0 (\xi - m_-)^{u_-} (m_+ - \xi)^{u_+} \l{ASSOLN} \; , \\
b(\xi) &=& b_0 (\xi - m_-)^{v_-} (m_+ - \xi)^{v_+} \l{BSSOLN} \; ,
\ea
where the constants $u_{\pm}$ and $v_{\pm}$ are
\ba
u_{\pm} &=& - {1\over 4} \left[1 \pm {n\xi_\alpha \over \Delta}\right]
 \nonumber \; , \\
v_{\pm} &=& \mp {\xi_\phi \over 2 \Delta}\nonumber \; .
\ea
Thus the qualitative behavior of the solutions is quite simple, approaching
power laws as $\xi\to m_{\pm}$ with the exponents $u_{\pm}$ and $v_{\pm}$.

Using Eqs.~(\r{EMET}) and~(\r{XI}) we see that the time ($dt$) in the string
frame is related to ($d\xi$) by
\ba
dt & = & a^3 b^{-n} e^{\phi} d\xi \quad , \\
 & \propto & (\xi - m_-)^{-1+w_-} (m_+ - \xi)^{-1+w_+} d\xi \; ,
\ea
where
\ba
w_{\pm} &=& 1 + 3u_{\pm} - nv_{\pm} + s_{\pm} \nonumber \; , \\
&=& -{1\over 4} \left[ 1 \pm {3n\xi_\alpha \over 4\Delta} \right] \; .
\l{WPM}
\ea
Although we cannot in general integrate this relation to obtain $t(\xi)$ in
closed form, we can solve for the time in the string frame in the limits
$\xi\to m_{\pm}$ to give
\be
t \propto | \xi - m_{\pm} |^{w_{\pm}} \l{PROPERT} \; .
\ee
We see that $\xi \rightarrow m_{\pm}$ in finite proper time if
$w_{\pm} >0$, which means that $\xi \rightarrow m_{+}$ in finite proper time if
$\xi_\alpha <0$, and  $\xi \rightarrow m_{-}$ in a proper time if
$\xi_\alpha >0$. Thus the proper time interval is always semi-infinite.
(The case $\xi_\alpha=0$ is excluded by the requirement that $\Delta\neq0$.)

{}From Eqs.~(\r{PROPERT}), (\r{ASSOLN}), (\r{BSSOLN}) and~(\r{PHISOLN})
we have the limiting behavior of the solutions as power laws
\ba
a(t) & \propto & |t-t_0|^{u_{\pm}/w_{\pm}} \; , \l{AOFT} \\
b(t) & \propto & |t-t_0|^{v_{\pm}/w_{\pm}} \; , \\
e^{\phi(t)} & \propto & |t-t_0|^{s_{\pm}/w_{\pm}} \; .
\ea
Thus we need to know the behavior of the ratio $u_{\pm}/w_{\pm}$ in order to
understand the late time behavior of the solutions for $a(t)$. Assuming
that we are dealing with events at late proper time as $\xi\to m_+$ and
$t\to\infty$ (i.e.~$\xi_\alpha>0$) then we have
\be
{u_+ \over w_+} = {n\xi_\alpha+\Delta\over
 3n\xi_\alpha+\Delta} \; , \nonumber
\ee
which is bounded by
\be
{1 \over 3} < {u_+ \over w_+} \leq {1 \over \sqrt3} \; .
\ee
The upper bound corresponds to the case $\xi_\phi=0$ where $b=$constant and
we recover the late-time behavior of the isotropic $D=4$ solutions.
This implies that in the string frame we do not obtain inflationary solutions
(by which we mean $d^2a/dt^2>0$) at {\it late} times. This of course does not
imply that we do not obtain inflationary solutions at earlier times. On the
contrary, all the solutions which initially contract
($da/dt<0$) but then expand at late times ($da/dt>0$) must undergo a period of
accelerated expansion.

Comparison with the isotropic $D=4$ solutions discussed in
Section~(\r{ssectISOhom}) is easiest if we use the conformally invariant
time coordinate defined by $d\eta=dt/a$, which is related to $\xi$ via
\be
\left| {\eta-\eta_0 \over \eta_*-\eta_0} \right| =
	\left( {\xi - m_- \over m_+ - \xi} \right)^{n\xi_\alpha/2\Delta} \; ,
\ee
where $\eta_*-\eta_0=a_0^2/n\xi_\alpha L$. Note that for $\xi_\alpha<0$ the
coordinate $\eta$ runs $-\infty$ to $\eta_0$ and for $\xi_\alpha>0$ we must
have $\eta_0\leq\eta<\infty$,
i.e.~the range for $\eta$ is always semi-infinite, coinciding with the
semi-infinite range for the proper time. To avoid any ambiguity
we introduce the non-negative variable $\tau$ defined in Eq.~(\ref{edeftau})
but here restricted to the spatially flat ($k=0$) case
\be
\tau = |\eta-\eta_0| \; ,
\ee
which runs from $\infty$ to $0$ for $\xi_\alpha<0$ and from $0$ to $\infty$ for
$\xi_\alpha>0$. We then have
\ba
a & = & {a_* \over \sqrt{2}} \left\{
	\left(\frac{\tau}{\tau_*}\right)^{1-\Delta/n\xi_\alpha}
		+ \left(\frac{\tau}{\tau_*}\right)^{1+\Delta/n\xi_\alpha}
	\right\}^{1/2} \; ,
\\
b & = & b_* \left(\frac{\tau}{\tau_*}\right)^{\xi_\phi/n\xi_\alpha} \; ,
\\
e^\phi & = & {e^{\phi_*} \over 2} \left\{
	\left(\frac{\tau}{\tau_*}\right)^{(n\xi_\phi-\Delta)/n\xi_\alpha}
	+ \left(\frac{\tau}{\tau_*}\right)^{(n\xi_\phi+\Delta)/n\xi_\alpha}
	\right\} \; .
\ea

Note that $b$ is a monotonic function of $\tau$.
However the qualitative behavior of $a$ depends on the value of
$n\xi_\alpha/\xi_\phi$ and can be separated into two cases.
\begin{enumerate}
\item $\Delta\leq|n\xi_\alpha|$ (requiring $|n\xi_\alpha|\leq|\xi_\phi|$).
$a(\tau)$ is monotonically increasing function;
\item $\Delta>|n\xi_\alpha|$ (requiring $|n\xi_\alpha|>|\xi_\phi|$). $a(\tau)$
contracts initially, ``bounces'' and then expands.
\end{enumerate}
The latter case includes the $D=4$ isotropic case (shown in Fig.~2)
where $\xi_\phi=0$ and $b$ remains constant. We then have
$\Delta/n\xi_\alpha=\sqrt{3}$ and we recover the solutions given in
Eqs.~(\r{ephitauhom}) \&~(\r{atauhom}) independent of $n$, the number
of extra dimensions while they remain static. Another example is when
we start with an isotropic contraction ($\dot{a}/a=\dot{b}/b<0$) at
$\tau=0$ as shown in Fig.~6. The size of the $n$-space continues to
decrease whereas the $H$-field prevents the 3-dimensional space
collapsing and it then expands at late times.

\vbox{\centerline{\epsfig{width=3.3in,file=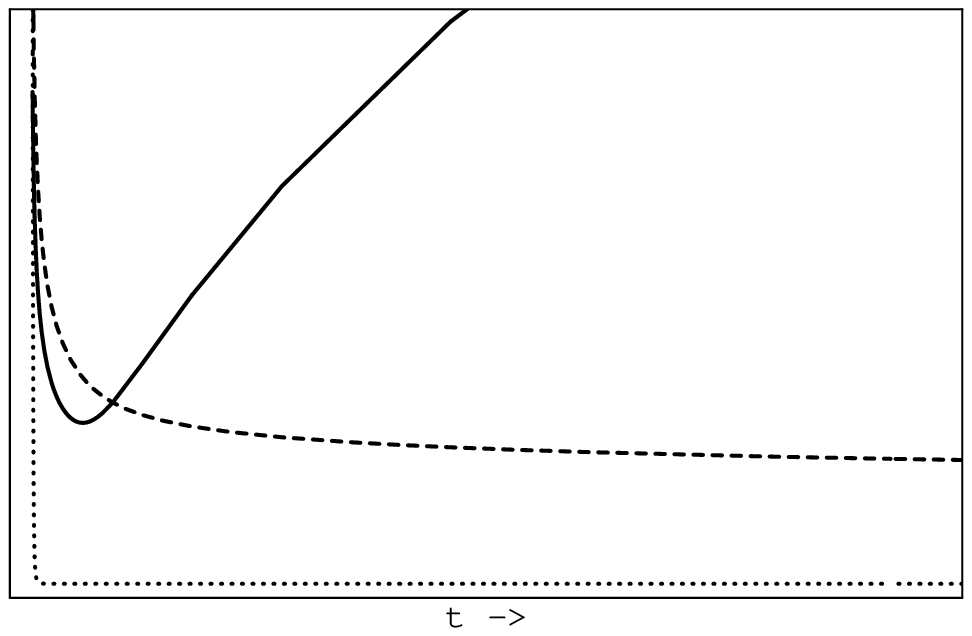}}
\noindent\small
{\bf Fig. 6:}
Scale factors $a$ (solid), $b$ (dashed) and dilaton $e^{\phi}$ (dotted) in
a  $D=4+6$ Bianchi I universe with homogeneous $h$,
starting from an approximately isotropic state at $t=0$.}

\bigskip

The critical case where $\Delta=|n\xi_\alpha|=|\xi_\phi|$ corresponds
to the case where the scale factor $a$ remains finite but non-zero at
$\tau=0$. Thus the curvature of the $4$-dimensional spacetime is
nonsingular, although it must be emphasized that both the dilaton and
$n$-space scale factor remain singular. This ``wormhole'' solution in
the $D=5$ case has been discussed recently by Behrndt and
F\"orste\cite{Beh+For 94}.

Although the $(4+n)$-dimensional dilaton $\phi$ displays the same range of
qualitative behavior as the scale factor $a$, the behavior of the
effective $4$-dimensional dilation $\varphi$ defined in Eq.(\ref{defvarphi})
is much more restrictive, evolving as
\be
e^\varphi = {e^{\phi_*} \over 2 b^n_*} \left\{
	\left(\frac{\tau}{\tau_*}\right)^{-\Delta/n\xi_\alpha}
	+ \left(\frac{\tau}{\tau_*}\right)^{\Delta/n\xi_\alpha}
	\right\} \; .
\ee
Thus it always has a minimum value at $\tau=\tau_*$.
As in the isotropic $D=4$ case, the presence of the homogeneous
antisymmetric tensor field introduces a minimum allowed value for the
effective gravitational coupling constant.

\subsection{Radiation solutions, $(\nabla h)^2=0$}

As in the $D=4$ isotropic case we can seek solutions corresponding to
the short wavelength limit of the Fourier decomposition of
$h(\eta,x^i)=h_q(\eta)\exp(iq_ix^i)$, where it is again convenient to
write the equation of motion for a Fourier mode $h_q$ using the
conformally invariant time coordinate, $\eta$, defined such that
$d\eta=dt/a=d\tt/\ta$. We then have in the extreme short wavelength
limit, $q\to\infty$,
\be
\left( \ta\tb^{n/2}e^{2\phi/(n+2)} h_q \right)''
 + q^2 \left( \ta\tb^{n/2}e^{2\phi/(n+2)} h_q \right) = 0
\; ,
\ee
and thus
\be
h(\eta,x^i) = m_q \frac{\tb^{n/2} \exp(-2\phi/(n+2))}{\ta} e^{iq_{\mu}x^{\mu}}
\; ,
\ee
where $m_q$ is a constant and $q_{\mu}$ is a null 4-vector restricted to the
4-dimensional metric $d\ts_4^2 = \ta^2(-d\eta^2+dx_i^2)$.

The corresponding energy-momentum tensor in the Einstein frame is
\be
\kappa_D^2 \tTH_{\mu}^{\nu} \propto {q_{\mu}q^{\nu} \over \ta^2\tb^n} \; ,
\ee
which when averaged over an isotropic distribution (with respect to the
3-space) gives the anisotropic (with respect to the whole $(n+3)$-space)
radiation fluid
\ba
- \kappa_D^2 \tTH_0^0 & \equiv & \kappa_D^2 \trho_H = {M \over \ta^4\tb^n}
\; , \\
\kappa_D^2 \langle \tTH_i^i \rangle & \equiv & \kappa_D^2 \tp_H
	= {1 \over 3} {M \over \ta^4\tb^n}
\quad {\rm (no\ sum)} \; , \\
\kappa_D^2 \tTH_J^J & \equiv & \kappa_D^2 \tq_H = 0 \quad {\rm (no\ sum)} \; .
\ea
The dilaton field again acts as an isotropic stiff fluid with
\be
\trho_\phi = \tp_\phi = \tq_\phi = \frac{\dot{\phi}^2}{2(n+2)\kappa_D^2}
\; ,
\ee
which does not appear in the Einstein evolution equations.

Substituting this $H$-field solution into the evolution Eqs.~(\r{eomphi}),
(\r{eomalphahom}) \&~(\r{eombetahom}) we have
\ba
\ddot{\phi} +\left( 3\dot{\alpha} + n\dot{\beta} \right) \dot{\phi}
	& = & 0 \; , \\
\ddot{\alpha} +\left( 3\dot{\alpha} + n\dot{\beta} \right) \dot{\alpha}
	& = & {M \over 3\ta^4\tb^n} \; , \\
\ddot{\beta} +\left( 3\dot{\alpha} + n\dot{\beta} \right) \dot{\beta}
	& = & 0 \; ,
\ea
while the constraint Eq.~(\r{aniconstraint}) becomes
\be
3\dot{\alpha}^2 + 3n\dot{\alpha}\dot{\beta} + {n(n-1) \over 2}\dot{\beta}^2 =
{\dot{\phi}^2 \over 2(n+2)} + {M \over \ta^4\tb^n} \; .
\l{EANICONSTRAINTRAD}
\ee

We can again obtain first integrals of all three evolution equations this time
using the conformally invariant time, $\eta$, so that
\ba
\phi' & = & {1 \over 3} {M \over \ta^2\tb^n} \ \eta_\phi \; , \\
\alpha' & = & {M \over \ta^2\tb^n} \ (\eta+\eta_\alpha) \; , \l{alphaprime}\\
\beta' & = & {M \over \ta^2 \tb^n} \ \eta_\beta \; , \l{betaprime}
\ea
where $\eta_\phi$, $\eta_\alpha$ and $\eta_\beta$ are constants of integration.

Using the same technique as in the homogeneous case, differentiating
Eq.(\r{alphaprime}) and substituting in for $\beta'$ using Eq.~(\r{betaprime})
we obtain a second order equation for $\alpha(\eta)$ whose first integral gives
\be
\alpha' = \frac{\eta + \eta_\alpha}
		{\eta^2 + (2\eta_\alpha+3n\eta_\beta)\eta + C} \; .
\l{alphaprimetoo}
\ee
The constant $C$ is given in terms of the other constants from the constraint
Eq.~(\ref{EANICONSTRAINTRAD}) as
\be
C = \eta_\alpha^2 + 3n\eta_\alpha\eta_\beta + {3n(n-1) \over 2}\eta_\beta
	- {3\over 2(n+2)}\eta_\phi^2 \; .
\ee
Comparing Eq.~(\r{alphaprimetoo}) with Eq.~(\r{alphaprime}) we see that
\be
\ta^2\tb^n = {M \over 3} \left( \eta^2 + (2\eta_\alpha+3n\eta_\beta)\eta
	+ C \right) \; ,
\ee
For each choice of integration constants we have two distinct
solutions. We have two semi-infinite intervals;
one where $\eta$ approaches $m_-$ from $-\infty$,
and the other for $\eta\geq m_+$ approaching $+\infty$. $m_{\pm}$ are
the roots of the above expression;
\be
m_{\pm} = -\eta_\alpha - {3\over2}n\eta_\beta
	\pm \Delta  \; ,
\ee
and
\be
\Delta = \half \sqrt{3n(n+2)\eta_\beta^2 + \frac{6}{n+2}\eta_\phi^2} \; .
\ee

Integrating the first-order equation for $\alpha(\eta)$,
Eq.(\ref{alphaprimetoo}), and similar expressions for $\beta$ and $\phi$
finally yields
\ba
\ta(\eta) & = & \ta_0 |\eta - m_-|^{p_-} |\eta - m_+|^{p_+} \; ,\\
\tb(\eta) & = & \tb_0 |\eta - m_-|^{q_-} |\eta - m_+|^{q_+} \; ,\\
e^{\phi(\eta)} & = & e^{\phi_0} |\eta - m_-|^{s_-} |\eta - m_+|^{s_+} \; ,
\ea
with the exponents
\ba
p_{\pm} & = & \half \left( 1 \mp {3n\eta_\beta \over 2\Delta} \right) \; ,\\
q_{\pm} & = & \pm {3\eta_\beta \over 2\Delta} \; , \\
s_{\pm} & = & \pm {3\eta_\phi \over 2\Delta} \; .
\ea

These are related to the solutions for the two scale factors in the original
string frame by the conformal transformation, Eq.~(\r{EMET}), which gives
\ba
a(\eta) & = & a_0 |\eta - m_-|^{u_-} |\eta - m_+|^{u_+} \; ,\\
b(\eta) & = & b_0 |\eta - m_-|^{v_-} |\eta - m_+|^{v_+} \; ,\\
\ea
with the exponents
\ba
u_{\pm} & = & \half \left( 1 \mp {3n\eta_\beta - \frac{6}{n+2} \eta_\phi
					\over 2\Delta} \right) \; ,\\
v_{\pm} & = & \pm {3\eta_\beta - \frac{3}{n+2} \eta_\phi \over 2\Delta}
\; .\\
\ea

As in the homogeneous case
the conformal time $\eta$ cannot in general be given in closed form
as a function of the proper time in the string frame, and can be given
instead only by the differential relation
$dt=ad\eta$. However, unlike the homogeneous case, we can identify a unique
late/early time behavior as $\eta\to\pm\infty$. We then find $dt\propto
\eta^{u_-+u_+} d\eta$ and thus $|t|\propto\eta^2$ giving
\ba
a & \propto & |\eta| \  \propto \ |t|^{1/2} \; ,\\
b, e^\phi & \to & {\rm constant} \; .
\ea
Thus the unique late time attractor for $\eta>m_+$ is the familiar solution
for an expanding $D=4$ radiation dominated universe. For $\eta<m_-$ the
asymptotic solution at early times represents the usual solution for a
collapsing radiation dominated universe. Both these solutions are singular
as $\eta\to m_{\pm}$ with $a\to0$ for $u_{\pm}>0$ or $a\to\infty$ for
$u_{\pm}<0$.

To compare these solutions with the isotropic case discussed in
Section~(\r{ssectISOrad}) we can write them in terms of the non-negative time
variable introduced for the spatially flat $D=4$ metric in Eq.~(\r{edeftau});
\be
\tau = \left\{
\begin{array}{cc}
m_- - \eta 	& {\rm for} \; \eta<m_- \; ,\\
\eta - m_+ 	& {\rm for} \; \eta>m_+ \; .
\end{array}
\right.
\ee
When $\eta\leq m_-$, $\tau$ decreases as the solutions approach the singularity
at $\tau=0$, while when $\eta\geq m_+$, $\tau$ increases away from the
singularity. Then $s(\eta)$, defined in Eq.~(\r{defs}), is given by
\be
s (\eta) = \frac{\tau(\eta)}{\tau(\eta) + \Delta} \; ,
\ee
and we can write
\ba
a & = & a_0 \tau^{1/2} (\tau + \Delta)^{1/2} s^{(2A-nB)/2} \; ,\\
b & = & b_0 s^{A+B} \; ,\\
e^\phi & = & e^{\phi_0} s^{(n+2)A} \; ,
\ea
where
\ba
A & = & {3\eta_\phi \over 2(n+2)\Delta} \; , \\
B & = & {3\eta_\beta \over 2\Delta} \; ,
\ea
so that $(n+2)(2A^2+nB^2)=3$. We can now identify the isotropic case with $b$
remaining constant and $A=-B=\pm\sqrt{3}/(n+2)$, which yields the results of
Section~(\r{ssectISOrad}).

\section{Conclusions}
\setcounter{equation}{0}

We have written down the general solutions to the low-energy equations
of motion derived from string theory for a four dimensional
Friedmann-Robertson-Walker cosmology including a homogeneous dilaton field
and antisymmetric tensor field.

In the absence of the anti-symmetric tensor field
($H_{\mu\nu\lambda}=0$) we recover the known vacuum results for the
evolution of the dilaton. In spatially flat models we have two
distinct branches; the decelerated branch where both the scale factor,
$a$, and the dilaton $\phi$ grow monotonically from zero at the
curvature singularity, and the accelerated branch where $a$ and $\phi$
decrease from infinity at the singularity.  These solutions are
invoked in the so-called ``pre-big-bang'' cosmologies based on the
accelerated branch expanding from $a$ and $\phi$ zero with low
curvature, and approaching the high curvature regime, where it must
evolve into the expanding decelerated branch with $a$ and
$\phi\to\infty$ as the curvature again vanishes.  But achieving such a
smooth transition in the context of the low energy effective theory
has proved difficult even with the inclusion of a dilaton
potential\cite{Bru+Ven 94}.

We have shown that the presence of the homogeneous $H$-field
introduces a minimum value for both the FRW scale factor and the
dilaton in $D=4$ models. Spatially flat models have $a\to\infty$ both
at the spacetime curvature singularity at $\eta=\eta_0$ and in the low
curvature limit ($\eta\to\pm\infty$). We no longer find two distinct
branches but rather a solution which smoothly interpolates between the
accelerated vacuum branch at high curvatures and the decelerated
branch at low curvatures. Our analytic solutions confirm some of the
results of the phase-plane analysis by Goldwirth and Perry
\cite{Gol+Per 93}.  Thus the extreme weak coupling limit, $\phi\to0$,
is attained only when $H_{\mu\nu\lambda}=0$. This is in contrast to
the results of \cite{Gas+Ven 93a} because our $H$-field does not obey
the requirement for $O(d,d)$ symmetry in their solutions in that our
$B_{\mu\nu}$ is not homogeneous. Rather, we consider solutions where
it is the field $H_{\mu\nu\lambda} \equiv
\partial_{[\mu}B_{\nu\lambda]}$ which is homogeneous. These solutions
are in fact related to the homogeneous vacuum solutions by an $SL(2,
R)$ duality transformation \cite{SL2R} where the dilaton and the
$H$-field are considered as the real and imaginary parts of a single
complex field\footnote{We are particularly grateful to N. Kaloper for
drawing this to our attention.}.

We find qualitatively similar solutions when we consider $4+n$
dimensions.  We give analytic solutions for a Bianchi I model with two
scale factors where the antisymmetric tensor field acts as a scalar
field on only three of the spatial dimensions. It tends to accelerate
their expansion with respect to the other $n$ dimensions producing an
anisotropic expansion even from isotropic initial conditions. The
effective four dimensional dilaton always has a minimum value. We
recover the isotropic $D=4$ solutions in the limit that the second
scale factor $b=$constant.

By treating the antisymmetric tensor field as a scalar field on the
four dimensional sub-space we can consider the short wavelength modes
which act like radiation as well as the long wavelength homogeneous
modes. We have also given analytic solutions in the case where this
radiation is isotropic on three spatial dimensions. The dynamical
effect of the $H$-field is then negligible near the curvature
singularity and we see both the accelerated and decelerated branches
seen in vacuum. However we find a unique late time attractor solution
in spatially flat models corresponding to the usual $D=4$ radiation
dominated solution in general relativity with $a\propto t^{1/2}$ and
$\phi\to$constant (and $b\to$constant in the two scale factor
anisotropic model). The dividing line between the short and long
wavelength regimes corresponds to wavelengths within or outside the
Hubble length.  Although this is a frame dependent quantity, for
power-law evolution, $a\propto |\eta-\eta_0|^p$, (which we find in all
asymptotic limits) any given wavelength must become ``long'' as
$\eta\to\eta_0$ and conversely any wavelength eventually becomes
``short'' as $|\eta|\to\infty$.

In models with non-zero spatial curvature (considered only in the isotropic
$D=4$ case) we find that although our solutions approach the flat solutions
near the spacetime curvature singularity they become dominated by the
spatial curvature as $\eta\to\pm\infty$ where the spacetime curvature
becomes small. This is hardly surprising if one considers the evolution of
the metric in the Einstein frame, of which we have made extensive use
throughout, where these string models correspond simply to a universe with
a massless scalar field (plus radiation for the short wavelength $H$-field)
and so there can be no inflation in this frame.

It is important to emphasize the limited cosmological era in which the
results presented here may be valid. While the appearance of a minimum
value for the scale factor of three dimensional space is intriguing, the
string metric is still in general singular at $\eta=\eta_0$. By solving
the field equations only
to lowest order in the string coupling constant $\alpha'$ we are
neglecting terms of order
$\alpha'R_{\mu\nu\lambda\kappa}R^{\mu\nu\lambda\kappa}$ with respect to
terms such as $R$, $H^2$, $(\nabla\phi)^2$, etc.\ in the field equations.
As all our solutions approach power-law evolution
($a_i\propto|\eta-\eta_0|^{p_i}$) as $\eta\to\eta_0$, then these terms
inevitably become divergent, so our solutions will only be good
approximations to the true evolution when the spacetime curvature is
sufficiently small. On the other hand we have neglected any potential
for the dilaton, assuming $V\ll(\nabla\phi)^2$. If we hope to recover
the standard hot big bang cosmology at late times then we will have to
include matter fields which presumably provide a potential to fix the
present day value of the dilaton (and thus satisfy observational limits on
the allowed
variation of the gravitational coupling strength within the solar system
today). Indeed if this potential for the dilaton or for any other fields
produces an inflationary era any memory of a preceding stringy era would
be all but erased. Neglecting the dilaton potential is only likely to be
valid while kinetic terms dominate. As we find $H^2\propto a^{-6}$ (for
homogeneous modes) this will only to be valid at sufficiently small $a$.

In summary, we have given general analytic solutions for the evolution of
an early, but sufficiently low-energy, stringy era
where the massless bosonic fields dominate the
dynamics. We have shown that the presence of the antisymmetric
tensor field has a dramatic effect on the evolution of the dilaton in four
dimensions and can also produce an anisotropic expansion in higher dimensional
models preferentially expanding three spatial dimensions.

\section*{Acknowledgements}

The authors are supported by the PPARC and would like to thank N.
Kaloper, A.~Liddle, A.~Tseytlin and K.~Behrndt for useful comments.
The figures were drawn with the help of {\it Mathematica}, \copyright
1988-93 Wolfram Research, Inc.


\nonfrenchspacing



\begin{thebibliography}{99}
\frenchspacing

\bibitem{effaction}
E. S. Fradkin \& A. A. Tseytlin, {\sl Nucl. Phys.} {\bf B261} 1 (1985);
C. G. Callan, D. Friedan, E. J. Martinez \& M. J. Perry, {\sl Nucl. Phys.}
{\bf B262} 593 (1985); C. Lovelace, {\sl Nucl. Phys.} {\bf B273} 413 (1985)

\bibitem{anton}
I. Antoniadis, C. Bachas, J. Ellis \& D.V. Nanopoulos,
{\sl Phys. Lett.} {\bf B221} 393 (1988);
{\sl Phys. Lett.} {\bf B257} 278 (1991);
{\sl Nucl. Phys.} {\bf B328} 117 (1989)

\bibitem{bailin}
D. Bailin \& A. Love, {\sl Phys. Lett.} {\bf B163} 135 (1985)

\bibitem{liddle 89}
A.R.Liddle, R.G.Moorhouse \& A.B.Henriques,
{\sl Nucl. Phys.} {\bf B311} 719 (1989)

\bibitem{ven 91}
G. Veneziano, {\sl Phys. Lett.} {\bf B265} 287 (1991)

\bibitem{campbell}
B.A. Campbell, N. Kaloper \& K.A. Olive,
{\sl Phys. Lett.} {\bf B277} 265 (19921)

\bibitem{tse+Vaf}
A. A. Tseytlin \& C. Vafa, {\sl Nucl. Phys.} {\bf B372} 443 (1992)

\bibitem{particular}
A. A. Tseytlin, {\sl Class. Quantum Grav.} {\bf 9} 979 (1992)

\bibitem{Tseytlin 92}
A. A. Tseytlin, {\sl Int. J. Mod. Phys.} {\bf D1} 223 (1992)

\bibitem{Gas+Ven 93b}
M. Gasperini \& G. Veneziano, {\sl Mod. Phys. Lett.} {\bf A8} 3701 (1993)

\bibitem{Gas+Ven 93a}
M. Gasperini \& G. Veneziano, {\sl Astroparticle Physics} {\bf 1} 317 (1993)

\bibitem{Gol+Per 93}
D. S. Goldwirth \& M. J. Perry, {\sl Phys. Rev. D}{\bf 49} 5019 (1994)

\bibitem{Bru+Ven 94}
R. Brustein \& G. Veneziano, {\sl Phys. Lett.} {\bf B329} 429 (1994)

\bibitem{MTW}
C. W. Misner, K. S. Thorne \& J. A. Wheeler, ``Gravitation'', W. H. Freeman \&
Company, New York (1973)

\bibitem{BDgravity}
C. H. Brans \& R. H. Dicke, {\sl Phys. Rev.} {\bf 124} 925 (1961)

\bibitem{Conformal}
R. H. Dicke, {\sl Phys. Rev.} {\bf 125} 2163 (1962)

\bibitem{Mimoso+Wands 94}
J. P. Mimoso \& D. Wands, {\sl Massless fields in scalar-tensor cosmologies}
preprint Sussex-Ast-94/4-1, gr-qc/9405025 (1994)

\bibitem{Mei+Ven 91}
K. A. Meissner \& G. Veneziano, {\sl Mod. Phys. Lett.} {\bf A6} 3397 (1991)

\bibitem{salam 82}
A. Salam \& J. Strathdee, {\sl Ann. of Phys.} {\bf 141} 316 (1982);
E.J. Copeland \& D.J. Toms, {\sl Nucl. Phys.} {\bf B255} 201 (1985)

\bibitem{Beh+For 94}
K. Behrndt \& S. F\"orste, {\sl String-Kaluza-Klein Cosmology} preprint
SLAC-PUB-6471, hep-th/9403179

\bibitem{SL2R}
A. Shapere, S. Trivedi \& F. Wilczek, {\sl Mod. Phys. Lett.} {\bf A6}
2677 (1991); A. Sen, {\sl Mod. Phys. Lett.} {\bf A8} 2023 (1993)

\end{thebibliography}
\end{document}